\documentclass[preprint,12pt]{elsarticle}
\usepackage{hyperref}

\usepackage{amssymb}
\usepackage{amsmath}
\DeclareMathOperator{\tr}{tr}

\begin{document}

\begin{frontmatter}



\title{Numerical simulations of non-relativistic stochastic fluids via the Metropolis algorithm}


\author[Bielefeld]{Mattis Harhoff\corref{cor1}} 
\ead{mharhoff@physik.uni-bielefeld.de}
\author[Bielefeld]{Sören Schlichting}
\author[Gießen,HFHF]{Lorenz von Smekal}

\cortext[cor1]{Corresponding author}

\affiliation[Bielefeld]{organization={Fakultät für Physik, Universität Bielefeld},
            addressline={Universitätsstr. 25},
            city={Bielefeld},
            postcode={33615},
            country={Germany}}
            
\affiliation[Gießen]{organization={Institut für Theoretische Physik, Justus-Liebig-Universität},
    addressline={Heinrich-Buff-Ring 16},
    postcode={35392},
    city={Gießen},
    country={Germany}}

\affiliation[HFHF]{organization={Helmholtz Forschungsakademie Hessen f\"ur FAIR (HFHF)},
    addressline={Campus Gießen},
    postcode={35392},
    city={Gießen},
    country={Germany}}

\begin{abstract}
Stochastic hydrodynamics provides a dynamical framework for the evolution of fluctuations in heavy-ion collisions, but poses significant challenges in numerical simulations, We present an algorithm for the simulation of non-relativistic stochastic fluids in two spatial dimensions in a box. We use the robust Metropolis algorithm, handling fluctuations and dissipation at once by systematically replacing dissipative terms in the hydrodynamic equations by random forces. The algorithm can easily be modified for numerical simulations of other hydrodynamic theories. We present test cases as well as numerical calculations of the renormalization of shear viscosity.
\end{abstract}

\begin{keyword}
Stochastic hydrodynamics



\end{keyword}

\end{frontmatter}



\section{Introduction}\label{sec:Intro}
Hydrodynamics is an effective description of many-body systems close to thermal equilibrium on large length and time scales, relying on an expansion of conserved fluxes in terms of local thermodynamic quantities and their gradients \cite{landau2013fluid}.
The fluctuation-dissipation theorem states that the non-reversible, dissipative fluxes arising in this gradient expansion are accompanied by stochastic thermal fluctuations \cite{landau1980statistical}. These can become large if the typical fluid cell is far from the thermodynamic limit, and in principle arbitrarily large and non-Gaussian near a critical point. In the case of non-linear hydrodynamic equations, fluctuations give both local dynamic contributions to dissipative transport as well as non-local contributions in the form of non-analytic effective terms in the gradient expansion \cite{Kovtun:2012rj}.\\
For these reasons, hydrodynamics with fluctuation, also called stochastic hydrodynamics, has received considerable attention in the heavy-ion community. A long-term goal is to include a systematic treatment of fluctuations into simulations of heavy-ion collisions, with stochastic hydrodynamics being one stage of the process. In such simulations, one can either use stochastic hydrodynamics directly, or choose a deterministic hydro-kinetic approach in which the evolution of the additional variables can be tuned by matching to stochastic hydrodynamic simulations. The underlying hydrodynamic theories are relativistic, requiring analysis and consistent treatment of hydrodynamic frames and their consequences for stability and causality in stochastic numerical simulations. An overview is given e.g. in the reviews by Bluhm et al. \cite{Bluhm:2020mpc} or Başar \cite{Basar:2024srd} and references therein.\\
In stochastic hydrodynamics as formulated by Landau and Lifshitz \cite{landau1980statistical, Landau:1957sat}, the deterministic hydrodynamic equations are promoted to stochastic partial differential equations (sPDEs) by inclusion of stochastic noise in the constitutive equations. However, numerical simulations of such Langevin-type theories have proven to be challenging \cite{Basar:2024srd}, since local noise leads to large gradients posing significant problems for numerical PDE solvers. In the continuum limit, the strength of the noise $\sim \delta^{(4)}(x-x')$ becomes even larger than background drift terms. Additionally, the choice of temporal discretization can survive the continuum limit, known as the multiplicative noise problem. Numerical implementations using the Langevin formulation exist both in non-relativistic and relativistic settings \cite{PhysRevE.76.016708, PhysRevLett.106.204501, doi:10.1137/120864520, Murase:2016rhl, Hirano:2018diu, Nahrgang:2017oqp, Bluhm:2018plm, Singh:2018dpk}, sometimes smoothing stochastic fluxes in order to avoid the aforementioned problems.\\
Recently, the Metropolis algorithm \cite{Metropolis:1953am, 10.1093/biomet/57.1.97} has been employed in numerical simulations of hydrodynamic theories \cite{Chattopadhyay:2025uqo, Chattopadhyay:2024bcv, Chattopadhyay:2024jlh, Basar:2024qxd, Bhambure:2024gnf}, introducing an additional accept-reject step when sampling stochastic noise. This reduces the infinite noise problem by likely rejecting large noise. The algorithm always generates a Markov chain following the correct equilibrium distribution, making simulations robust and free of the multiplicative noise issue. Stability issues of the underlying hydrodynamic theories can be avoided, since dissipative terms are treated independently from the ideal advection dynamics.\\
As a first step towards more complicated theories, we present in this work an algorithm for the numerical simulation of non-relativistic stochastic fluids in two spatial dimensions in a box. In Section \ref{sec:Theory}, we define our model by writing down equations of motion and defining the stochastic noise. Section \ref{sec:Numerics} explains the implementation of the algorithm consisting of independent schemes for ideal and stochastic dissipative dynamics. Section \ref{sec:Tests} discusses tests of the implementation, followed by Section \ref{sec:Renorm} in which the algorithm is applied to study the renormalization of transport coefficients numerically. We present our conclusions in Section \ref{sec:Concl}.

\section{Model}\label{sec:Theory}
\subsection{Equations of motion}
The equations of motion of non-relativistic hydrodynamics are conservation laws for the particle number density $n$, total energy density $e$ and momentum density $\vec{\pi}$ and their associated fluxes:
\begin{subequations}\label{eq:conservation_laws}
    \begin{align}
        \partial_{t}{n}+\partial_{i}J^{i}_{n}&=0 \\
        \partial_{t}{e}+\partial_{i}J^{i}_{e}&=0 \\
        \partial_{t} \pi^{j}+\partial_{i}J^{i}_{\pi^j}&=0
    \end{align}
\end{subequations} We define the local rest-frame (LRF) of the fluid as comoving with the particle flux, such that $\vec{J}_n|_\text{LRF}=0$. The fluxes in the general frame are given by the fluxes in the LRF plus Galilean boost terms:
\begin{subequations}\label{eq:fluxes}
    \begin{align}
        J_n^{i} &= \left. J_n^i \right|_\text{LRF}+ n v^i  \\
        J_e^{i} &= \left. J_e^i \right|_\text{LRF} + e v^{i} + \left. J_\pi^{ij} \right|_\text{LRF} v^{j} \\
        J_{\pi^j}^{i}\equiv J_\pi^{ij} &= \left. J_\pi^{ij} \right|_\text{LRF} + m n v^{i} v^{j}
    \end{align}
\end{subequations}
The local fluid velocity is given by $\vec{v}=\vec{\pi}/mn$, with $m$ being the mass of a constituent particle. Following the prescription by Landau and Lifshitz \cite{landau1980statistical, Landau:1957sat}, we introduce thermal fluctuations in the LRF as
\begin{subequations}\label{eq:stoch_fluxes}
    \begin{align}
        \left. J_e^i \right|_\text{LRF} &= \xi_\kappa^i, \\
        \left. J_\pi^{ij} \right|_\text{LRF} &= (p+\xi_\zeta)\delta^{ij} + \frac{1}{2} \left( \Xi_\eta^{ij}+\Xi_\eta^{ji}-\tr(\Xi_\eta)\delta^{ij} \right),
    \end{align}
\end{subequations}
with $\xi^i_\kappa$, $\xi_\zeta$, $\Xi_\eta^{ij}$ being stochastic noise terms. Note that in the LRF, the particle flux has no correction by definition, and that the pressure $p$ is an isotropic equilibrium contribution to the momentum flux tensor. The stochastic momentum fluxes are introduced such that they respect the tensor structure of $J_\pi^{ij}$, with an isotropic bulk noise and a traceless, symmetric shear noise. The stochastic terms give rise to non-equilibrium fluxes, which should on average reproduce the first-order viscous fluxes
\begin{subequations}\label{eq:avg_fluxes}
    \begin{align}
        \langle\!\langle \left. J_e^i \right|_\text{LRF}\rangle\!\rangle &= L_\kappa \;\partial^i \left(\frac{1}{T}\right), \\
        \langle\!\langle \left. J_\pi^{ij} \right|_\text{LRF}\rangle\!\rangle &= \left[p-\frac{L_\zeta}{T} \big(\Vec{\nabla}\cdot\Vec{v}\big)\right]\delta^{ij}-\frac{L_\eta}{T} \sigma^{ij}
    \end{align}
\end{subequations}
to linear order in the LRF, where by $\langle\!\langle\cdot\rangle\!\rangle$ we denote an average over all realizations of the fluctuations, $T$ is the local temperature and $\sigma^{ij}=\frac{1}{2}[ \partial^j v^i+\partial^i v^j-\delta^{ij}(\vec{\nabla}\!\cdot\!\vec{v})]$ the shear stress tensor in two dimensions. Instead of heat conductivity $\kappa$, bulk viscosity $\zeta$ and shear viscosity $\eta$, we use constant linear transport coefficients $L_\kappa, L_\zeta, L_\eta$ such that
\begin{align}
    \kappa=\frac{L_\kappa}{T^2},\qquad
    \zeta=\frac{L_\zeta}{T},\qquad
    \eta=\frac{L_\eta}{2T}.
\end{align}
For the moment, we will not fix the form of the noise terms, but note that they are local and their strength is related to the transport coefficients via fluctuation-dissipation relations. Moreover, they take a form such that they drive the system to its microcanonical equilibrium distribution
\begin{align}\label{eq:p_eq}
    p_{N_0, E_0, \vec{P}_0}[n,e,\vec{\pi}]=\frac{1}{Z}e^{S[n,e,\vec{\pi}]}\delta(N[n]-N_0)\delta(E[e]-E_0)\delta(\vec{P}[\vec{\pi}]-\vec{P}_0).
\end{align}
Here, $S[n,e,\vec{\pi}]$ denotes the total entropy and thus the logarithm of the statistical weight of a configuration. The delta functions ensure overall conservation of the conserved quantities, with $N, E, \vec{P}$ being the total particle number, energy and momentum of a configuration and $N_0, E_0, \vec{P}_0$ being control parameters. $Z$ denotes the partition sum.

\subsection{Equations of state}
In order for the equations in (\ref{eq:conservation_laws}) to form a closed system, an equation of state is needed. For simplicity, we work with an ideal gas $p=nT$, where (again in two dimensions) the temperature can be calculated from the internal energy density $u\equiv e-\frac{\vec{\pi}^2}{2mn}$ as $u=nT$, such that the pressure is given by $p(n,e,\vec{\pi})=u(n,e,\vec{\pi})$. We emphasize that the equation of state is an input to the algorithm, which is independent of this assumption. The entropy density of the ideal gas in two dimensions reads $s=n\left( 2+\log\frac{um}{2\pi n^2}\right)$.

\subsection{Units}
All appearing units can be expressed in terms of temperature, mass, length and time. We fix the units by setting $\hbar=1$ and $k_B=1$, as well as $m=1$ and lattice spacing $\Delta x=1$. All units are to be understood in these lattice units. The dimensionless parameters of the theory include the usual dimensionless parameters of non-relativistic hydrodynamics as well as fluctuation-related ones following from the fluctuation-dissipation relations.

\section{Algorithm}\label{sec:Numerics}
We discretize the system on a two-dimensional square $L\times L$ lattice with periodic boundary conditions and lattice spacing $\Delta x$, such that the total volume is $V=(L \Delta x)^2$. In the following, we use only finite-volume schemes, conserving the primitive variables exactly up to numerical precision.\\
We split a numerical time evolution step $t\to t+\Delta t$ into two substeps, evolving first only the terms governing the ideal non-dissipative advection dynamics and then fluctuations and dissipation. In practice, we choose $\Delta t$ by using an adaptive integrator for the ideal step. However, since the ideal and dissipative dynamics have different time scales, it can in general be advised to split one of the two substeps into many proportionally smaller ones.

\subsection{Ideal hydrodynamics}
\begin{figure}[htb]
    \centering
    \includegraphics[width=0.8\textwidth]{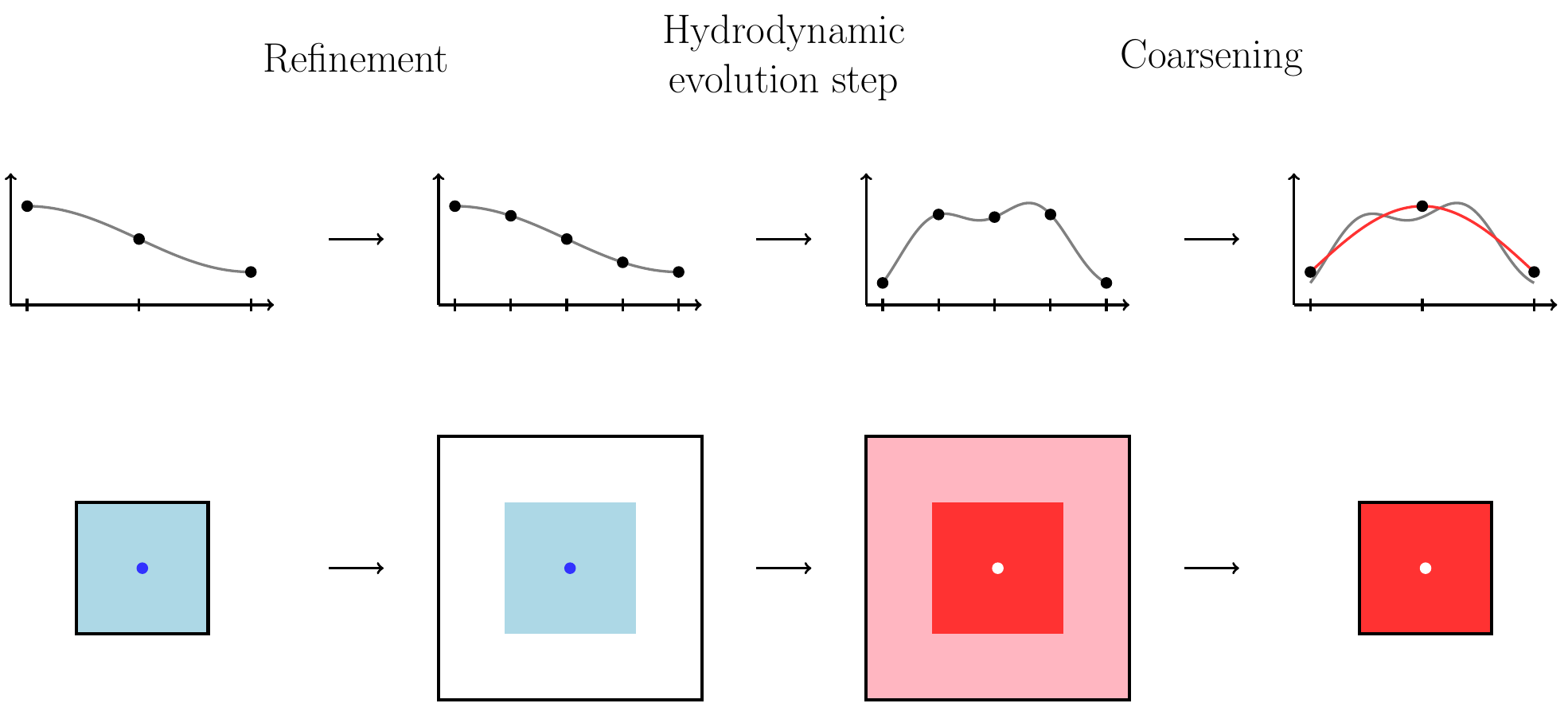}
    \caption{Mesh refinement via zero-padding Fourier interpolation in real (top) and Fourier space (bottom). The finer lattice has the same volume, such that its Fourier spectrum has additional modes only in the padding around the original spectrum. The refinement is effectively a trigonometric interpolation, copying the spectrum and setting modes in the padding to zero. After the hydrodynamic evolution step, there can be small occupations of the fast modes in the padding. These are removed in the coarsening step, such that the coarse configuration is given by the slow envelope of the fine configuration.}
    \label{fig:refinement}
\end{figure}

We solve the ideal hydrodynamic equations employing the semi-discrete Kurganov-Tadmor (KT) scheme \cite{Kurganov:2000ovy} together with a fifth-order Dormand-Prince integration with adaptive step size as formulated in \cite{Hairer}.\\
Since the ideal dynamics do not feature dissipation, one would like the ideal substep to conserve the probability functional of fluctuations in (\ref{eq:p_eq}) exactly. However, the chain rules linking entropy conservation to conservation of the primitive variables are not fulfilled at the discrete level, meaning that the scheme conserves entropy only approximately. While schemes explicitly enforcing these chain rules can sometimes be constructed \cite{Chattopadhyay:2024bcv, Morinishi1998}, this is not possible in the general case. In the case of the KT scheme, numerical dissipation in regions of large gradients stems mainly from flux reconstruction via flux limiters and local maximum propagation speeds. This effect is in principle wanted in order to avoid so-called spurious oscillations \cite{Kurganov:2000ovy}.\\
Nevertheless, the choice of gradient discretization becomes negligible in the case of configurations that are smooth on scales $\sim\Delta x$, meaning that one would ideally like to take a continuum limit $V=\text{const.}$, $\Delta x\to0$ while keeping fluctuations at a fixed characteristic scale $\gg \Delta x$. While one could realize this limit by spatially smoothing the stochastic fluxes, we avoid such a procedure here, since it would make the Metropolis update non-local. Instead, we consider the coarse Metropolis lattice to be physical and perform a zero-padding Fourier mesh refinement before carrying out the ideal update step on the finer lattice, coarsening afterwards by removing the additional fast Fourier modes in the padding. The procedure is visualized in Fig.~\ref{fig:refinement}.

\subsection{Stochastic step}
\begin{figure}[h]
    \centering
    \begin{minipage}{0.49\textwidth}
        \centering
        \includegraphics[height=4cm]{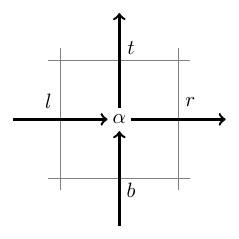}
    \end{minipage}
    \hfill
    \begin{minipage}{0.49\textwidth}
        \centering
        \includegraphics[height=3.5cm]{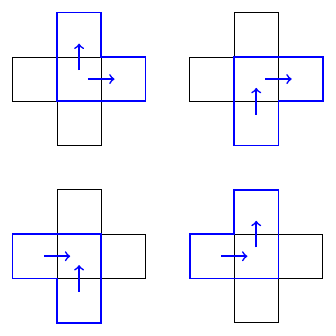}
    \end{minipage}
    \caption{Left: Fluxes transporting the conserved quantities in and out of cell $\alpha$ with faces $q_\alpha\in\{l,r,t,b\}$. Right: Possible orientations of L-update.}
    \label{fig:cell}
\end{figure}

The stochastic step consists of an Euler integration step of the stochastic fluxes given in Eqs.~(\ref{eq:stoch_fluxes}), sampled on the interfaces between the numerical cells. For example, the total energy change in cell $\alpha$ due to fluxes on the faces $q_\alpha$ with outwards-facing normal vectors $\vec{\mathfrak{n}}_{q_\alpha}$ is $\Delta E_\alpha = - \sum_{q_\alpha} J_{e,q_\alpha}^j \mathfrak{n}_{q_\alpha}^j \Delta x \Delta t$ with an equivalent equation for momentum. The process is schematically depicted on the left side of Fig.~\ref{fig:cell}.
Using these expressions, one can approximate the total change in entropy $S=\sum_\alpha S_\alpha$ to lowest order as function of the stochastic noise terms on all cell faces. The expression $\Delta S/\Delta t$ in the continuum limit $\Delta x, \Delta t\to 0$ only matches the thermodynamic continuum prescription $\partial_t S$ if one correlates one horizontal cell face with one vertical one in each cell, e.g.~by setting $\Xi^{ij}_{\eta, \alpha, t}=\Xi^{ij}_{\eta, \alpha, r}$ and analogously for the other stochastic terms. This choice plays a crucial role in the case of momentum, since $J_\pi^{ij}=J_\pi^{ji}$ is symmetric in first-order viscous hydrodynamics. Hence, terms containing $\partial_i v_j$ only have correct identical prefactors in $\partial_t S$ if the noise terms are identical for both directions. The geometry of the Metropolis update should in other words respect the symmetries of the underlying equations, here corresponding to the conservation of angular momentum.\\
Based on this argument, we perform each local Metropolis update by selecting a cell and two of its neighbors in an L-shape. We stack the updates by tiling the plane with crosses and in each cross performing Metropolis updates using all four possible L-orientations (as depicted in the right panel of Fig.~\ref{fig:cell}), considering all tilings such that each cell has been the center of such a cross once. Using this procedure, each cell face will be updated four times, such that we perform each microscopic update using a time step $\frac{1}{4}\Delta t$.\\
In a microscopic update, identical fluxes are proposed on the two faces. The fluxes are constructed from $\xi_\kappa^i$, $\xi_\zeta$ and $\Xi_\eta^{ij}$ drawn from uncorrelated normal distributions with mean zero. The proposal is accepted depending on the total change of entropy with probability $\min(1, e^{\Delta S})$.\\
To lowest order, one can write $\Delta S=\vec{g}\cdot\vec{\xi}$, where $\vec{g} $ is a vector proportional to $ \Delta t (\Delta x)^2$ containing gradients, and $\vec{\xi}$ is a vector containing the stochastic fluxes, with the components going over all noises and corresponding gradient factors in the updated cell. The variances $\sigma_i^2$ are tuned so that the average Metropolis fluxes reproduce the expectation values given in Eqs.~(\ref{eq:avg_fluxes}). Using the above approximation, it reads
\begin{subequations}\label{eq:fluxavg}
       \begin{align}
        \langle\!\langle\xi_k\rangle\!\rangle &\approx \int \left(\prod_{i=1}^N \frac{d\xi_i}{\sqrt{2\pi\sigma_i^2}}\exp\left[ -\frac{\xi_i^2}{2\sigma_i^2} \right]\right) \cdot \min(1, e^{\Vec{g}\cdot\Vec{\xi}}) \cdot \xi_k \\
        &=\frac{\sigma_k^2}{2}g_k \Big( 1 + \mathcal{O}\big( \textstyle\sum_i g_i^2\sigma_i^2 \big)\Big).
    \end{align}
\end{subequations}
Tuning to match Eqs.~(\ref{eq:avg_fluxes}), we obtain
\begin{align}
    \sigma_\kappa^2 = \frac{2 L_\kappa}{(\Delta x)^2 \; \Delta t},
    \qquad\quad
    \sigma_\zeta^2=\frac{2L_\zeta}{(\Delta x)^2 \; \Delta t},
    \qquad\quad
    \sigma_\eta^2=\frac{2L_\eta}{(\Delta x)^2 \; \Delta t}.
\end{align}
The factors of $\Delta t$ and $\Delta x$ can be interpreted as coming from continuous fluctuation-dissipation relations containing $\delta^{(2)}(\vec{x}-\vec{y})\delta(t-t^\prime) \sim 1/[(\Delta x)^2 \; \Delta t]$. The fact that the fluctuation-dissipation relations are otherwise identical to the continuum ones is due to $\Delta S\approx \partial_t S \Delta t $ vanishing as $\Delta t\to 0$, such that changes are always accepted and Langevin dynamics recovered, from which the original fluctuation-dissipation relations are derived.

\section{Numerical simulations}\label{sec:Tests}
We test the algorithm by testing both, the correct microscopic and macroscopic equilibration of the system. Macroscopically, by recording time dependent histograms of the conserved densities over all lattice site, one can verify that the system converges to the correct equilibrium distribution in (\ref{eq:p_eq}) over time.\\
Microscopically, the non-trivial generation of the constitutive relations in (\ref{eq:avg_fluxes}) can be tested by recording stochastic fluxes generated by the algorithm as function of the associated gradient term on the right-hand sides of Eqs.~(\ref{eq:avg_fluxes}). One can then numerically take an average over all realizations of the fluctuations by taking an average in bins of the gradient.
\begin{figure}[h]
    \centering
    \includegraphics[width=0.8\textwidth]{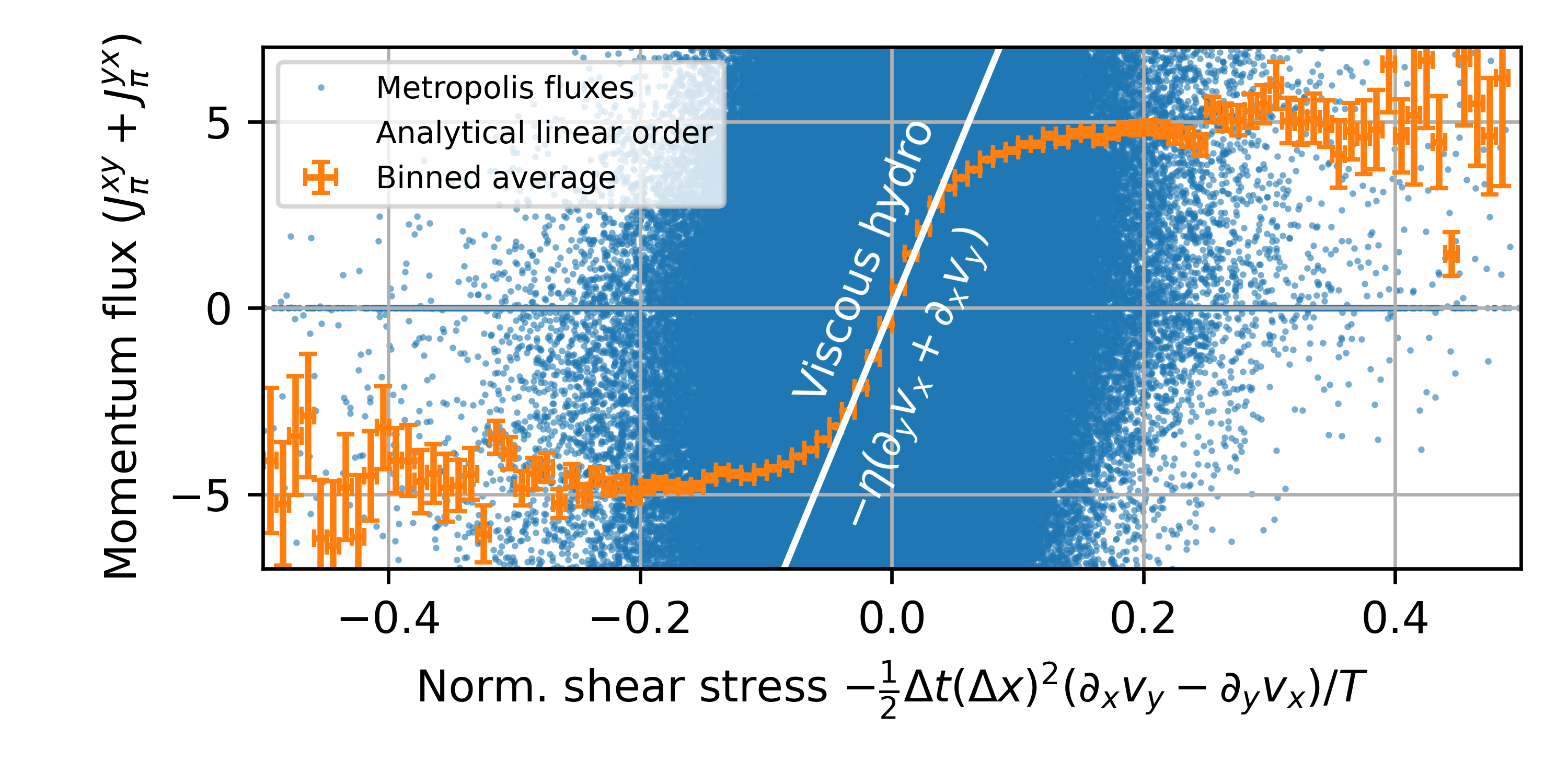}
    \caption{Averaged Metropolis fluxes recovering viscous hydrodynamics to first order in the shear channel.}
    \label{fig:const_rel}
\end{figure}
Fig.~\ref{fig:const_rel} shows the results of this analysis in the case of shear stress, with blue points denoting the individual fluxes generated by the Metropolis algorithm and the binned average given in orange. For small shear stress, the Metropolis algorithm recovers on average the prescription by first-order viscous hydrodynamics to linear order, as predicted by Eq.~(\ref{eq:fluxavg}). Since the Metropolis step is likely to reject unphysically large changes, for higher gradients the average flux saturates. Note that the scaling of the $x$-axis depends on the scale of the gradients prescribed by fluctuation-dissipation relations as well as the values of $\Delta x$ and $\Delta t$. These parameters were here deliberately chosen to cover a large range of shear stresses and are not necessarily part of a physical parameter space. An analogous analysis was performed for heat conduction and the sound channel.

\section{Renormalization of transport coefficients}\label{sec:Renorm}
\begin{figure}[h]
    \centering
    \includegraphics[width=0.6\textwidth]{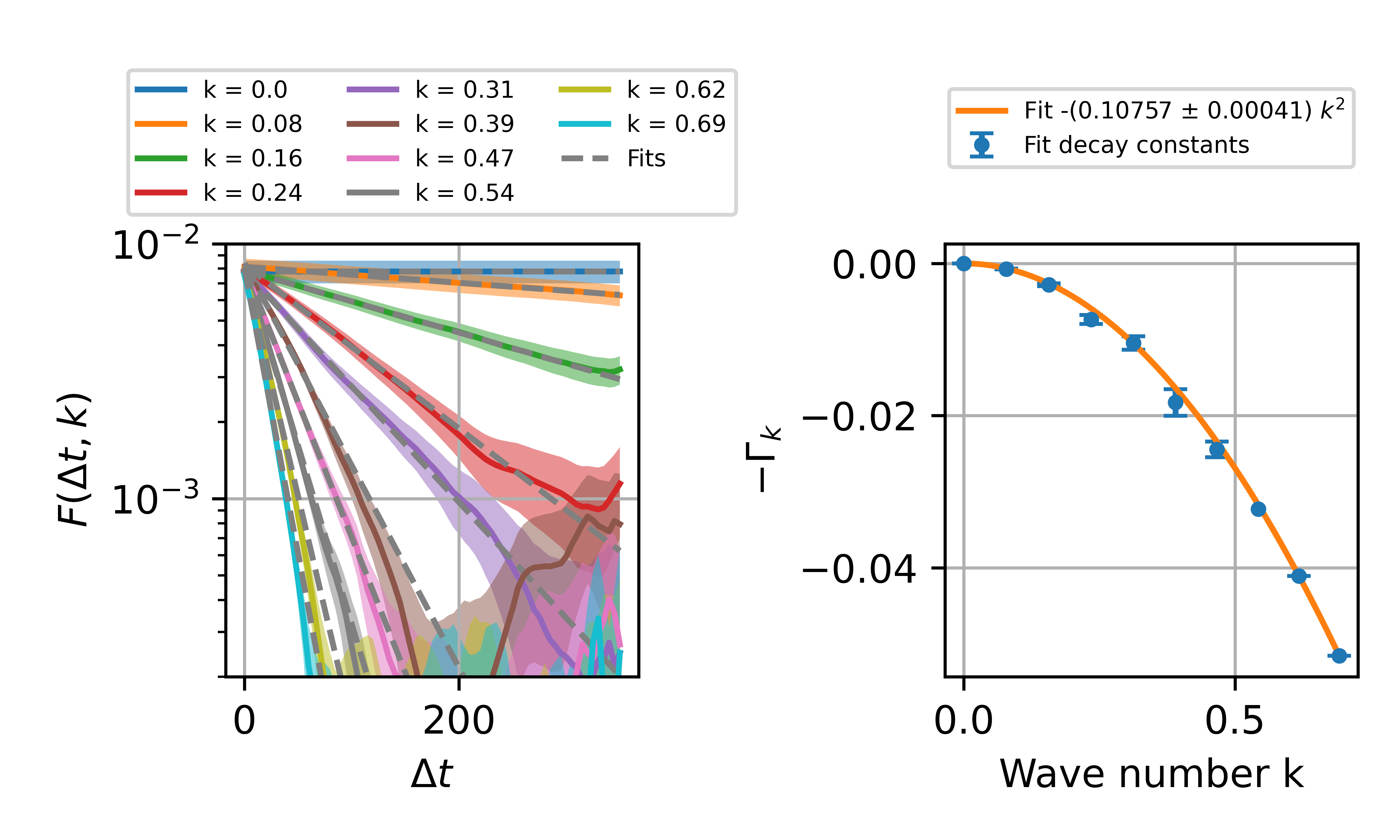}
    \includegraphics[width=0.27\textwidth]{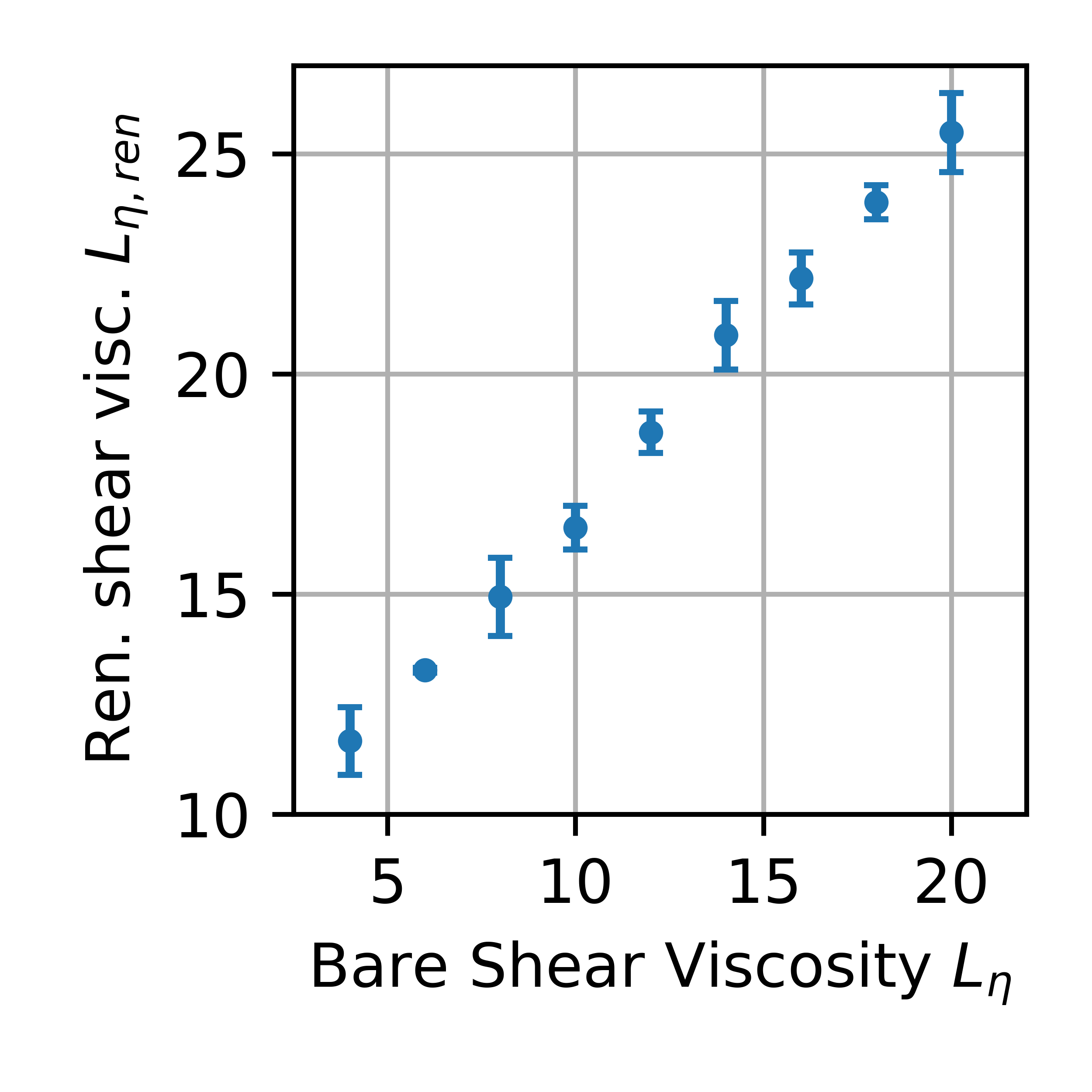}
    \caption{Left: Numerical shear-shear correlators $F(\Delta t, k)$ for $L=80$, $\Delta x=1$, $L_\eta=6$ and $L_\kappa=L_\zeta=0$ in grand-canonical ensemble with $\beta=0.4$, $\mu=10.0$. Middle: Parabolic fit to decay constants $\Gamma_k$ extracted from fits in left panel. Right: Renormalization of shear viscosity. $L_{\eta, \text{ren}}$ was extracted from shear-shear correlators, while $L_\eta$ denotes the microscopic bare value. Since results in the given region agreed for different mesh refinement factors, no mesh refinement was used for these calculations.}
    \label{fig:ren_shear}
\end{figure}

Fluctuations together with interactions of the hydrodynamic modes change the dissipative dynamics of the fluid both locally and non-locally \cite{Basar:2024srd, Kovtun:2011np}. One non-trivial effect is the renormalization of transport coefficients and the equation of state from contributions of collective excitations. It has for example been shown that contributions of sound modes lead to a non-zero lower limit of the shear viscosity in relativistic fluids \cite{Kovtun:2011np}.\\
In this section, we demonstrate how to extract effective transport coefficients from numerical simulations in order to compare to the bare microscopic input parameters. For example, shear modes with wave vector $\vec{k}$ decay exponentially in the linearized theory at a rate $\Gamma_k=\frac{L_\eta}{2mu_0}k^2$, where $u_0$ is the internal energy density in equilibrium.
The thermal shear-shear correlator is then given by
\begin{align}\label{eq:shearshearcorr}
    F(\Delta t, k)\equiv\langle\vec{\pi}^{\perp*}_k(t_0+\Delta t)\vec{\pi}^\perp_k(t_0)\rangle
    =\langle|\delta\vec{\pi}_k^\perp|^2\rangle e^{-\Gamma_k \Delta t},
\end{align}
in which $\langle\cdot\rangle$ denotes a thermal average that we realize numerically by averaging both over $t_0$ and initial conditions.  In the full non-linear and interacting theory, the shear correlator carries loop corrections with an exponential decay at a rate $\propto L_{\eta, \text{ren}}k^2$ being only the leading contribution. Numerically, we take e.g. $\vec{k} \parallel \hat{x}$ and $\vec{\pi}^\perp \parallel \hat{y}$ and consider only the lowest few $\vec{k}$ on the lattice, taking the physical wavenumber to be the eigenvalue of the central difference operator. From the time evolution of these modes on the lattice, one can extract the shear-shear correlator in equilibrium as given by the left-hand side of Eq.~(\ref{eq:shearshearcorr}). This shows to good approximation (for small $\Delta t$) an exponential decay, such that $\Gamma_k$ can be extracted by fitting a straight line to the logarithm of the correlator. Taking for small $k$ the mean of the decay constant divided by $k^2$ gives a numerical value of the renormalized shear viscosity $L_{\eta, \text{ren}}$.\\
Fig. \ref{fig:ren_shear} displays an example of the extraction procedure in the left and middle panels and on the right an example of $L_{\eta, \text{ren}}$ as a function of the bare $L_{\eta}$. 
The data follows to a good approximation a straight line with slope $\sim 1$ and non-zero offset. Numerically, both, the regions of small and large $L_{\eta}$ are problematic. For large shear viscosity, fluctuations and thus gradients are large, posing a problem for the numerical ideal hydrodynamic solver. In practice, we see that the Metropolis algorithm can correct for some numerical dissipation by adjusting the acceptance rate. On the other hand, this is not possible in the case of small fluctuations, where data might be dominated by numerical dissipation. The problem could be mitigated by using a multistep scheme, in which one first performs multiple ideal steps before taking a stochastic step. A systematic study of such effects could be performed by comparing numerical with analytical results.

\section{Conclusion}\label{sec:Concl}
In these proceedings, we have demonstrated how to construct and test an algorithm for numerical simulations of non-relativistic stochastic fluids using the Metropolis algorithm. Our method can easily be applied in other spatial dimensions and to other hydrodynamic theories or sPDEs. While we found that the numerical dynamics in certain parameter ranges are strongly influenced by the performance of the solver of the ideal hydrodynamic equations on noisy configurations, we propose to perform a mesh refinement before the ideal step, such that the performance of the ideal solver can be improved in a systematic manner. This is in contradistinction to other works, which instead smooth the stochastic fluxes. Moreover, our method does not depend on the specific implementation of the ideal step, such that one can in principle use schemes that explicitly conserve entropy or rather the probability functional of fluctuations, if available. We have demonstrated how to extract transport coefficients and how to study the renormalization of these coefficients numerically.\\
One could extend the algorithm to relativistic hydrodynamics. The density frame \cite{Bhambure:2024gnf} naturally lends itself due to dissipative fluxes being spatial, leading to a notion of global time that is necessary for Metropolis simulations. Even in its present formulation, it can be used to study the rich physics of stochastic fluids: For example, one could study full critical dynamics including sound modes by using a critical equation of state.

\section*{Acknowledgements}
We thank Frederic Klette, Dirk Rischke, Johannes V. Roth, Thomas Schäfer, Leon J. Sieke, and Derek Teaney for fruitful discussions. This work was supported by the Deutsche Forschungsgemeinschaft (DFG, German Research Foundation) through the CRC-TR 211 ‘Strong-interaction matter under extreme conditions’ – project no.\ 315477589 – TRR 211. The authors gratefully acknowledge the funding of this project by computing time provided by the Paderborn Center for Parallel Computing (PC2).

%
%
\bibliography{refs}

\end{document}